\documentclass[3p,times,procedia]{elsarticle}
\usepackage{nupha_ecrc}
\usepackage{lineno}

\volume{00}

\firstpage{1}

\journalname{Nuclear Physics A}

\runauth{Leticia Cunqueiro}

\jid{nupha}

\jnltitlelogo{Nuclear Physics A}

\usepackage{amssymb}

\usepackage[figuresright]{rotating}

\begin{document}

\begin{frontmatter}



\dochead{}

\title{Jet shapes in pp and Pb--Pb collisions at ALICE}



\author{Leticia Cunqueiro for the ALICE Collaboration}

\address{University of M\"unster and CERN}

\begin{abstract}
 The aim of this work is to explore possible medium modifications to the
 substructure of inclusive charged jets in Pb-Pb relative
 to proton-proton collisions by measuring a set of jet shapes. The set of
 shapes includes the radial moment, $g$, and the momentum dispersion
 $p_{\mathrm{T}}$D. They provide complementary information on the
 fragmentation and can help to discriminate between two different
 scenarios: intra-jet broadening or collimation as a result of jet
 quenching. 
\noindent The shapes are measured in Pb--Pb collisions at
$\ensuremath{\sqrt{s_{\mathrm{NN}}}}$ = 2.76 TeV with a constituent cutoff of 0.15 GeV/c and jet
 resolution $R=$ 0.2. New techniques for background subtraction are
 applied and a two-dimensional unfolding is performed  to correct the
 shapes to particle level. The corrected jet shapes for jet $p_{\mathrm{T}}$ $40\le
 p_{\mathrm{T,jet}} \le 60$ GeV/c are presented and discussed. The
 observed jet shape modifications suggest that the in-medium fragmentation is
 harder and more collimated than vacuum fragmentation as obtained by a
 PYTHIA calculation. The PYTHIA calculation is validated with proton-proton data
 at 7 TeV.

\end{abstract}


\end{frontmatter}


\section{Jet shapes}
\label{intro}
 \noindent  The hot and dense medium created in Heavy Ion Collisions is expected to modify
 the jet yield and fragmentation relative to pp collisions. The measurement of
 such modifications gives insight into the mechanisms of energy loss
 of partons in the medium and ultimately into the
 properties of the medium itself. The aim of this work is to characterize changes
 in the intrajet distribution using observables that are well-defined,
 preserving the infrared and collinear safety of the measurement and thus allowing for
 a direct connection to the theory. 
 \noindent In this analysis we focus on two jet shape observables that probe complementary
aspects of the jet fragmentation, namely the first radial moment $g$
 and the momentum dispersion $p_{\mathrm{T}}$D \cite{Larkoski:2014pca}. 

 \noindent The radial moment $g$ is defined as: 
\begin{equation}
\label{eq:angu}
g = \sum_{i \in jet} \frac{p_{\mathrm{T}}^{\mathrm{i}}}{p_{\mathrm{T,jet}}}  \vert \Delta R_{\mathrm{i,jet}} \vert 
\end{equation} 
where $p_{\mathrm{T}}^{\mathrm{i}}$ stands for the momentum of constituent $i$ and
$\Delta R_{\mathrm{i,jet}}$ is the distance in $\eta$, $\phi$ space between
constituent i and the jet axis. This shape measures the radial
energy profile of the jet. As an illustrative example, gluon jets
fragment more and thus are broader and have higher
radial moment than quark jets \cite{harvard}.

\noindent The  momentum dispersion $p_{\mathrm{T}}$D is defined as: 

\begin{equation}
\label{eq:angu}
 p_{T}D = \frac{\sqrt{\sum_{i \in jet} p_{\mathrm{T,i}}^{2}}}{\sum_{i \in jet} p_{\mathrm{T,i}}}.
\end{equation} 
This shape measures the second moment of the constituent p$_{T}$
distribution in the jet and tells how hard/soft the fragmentation is. For
example, in the extreme case of few constituents carrying a
large fraction of the jet momentum, $p_{\mathrm{T}}D \rightarrow$ 1, while in
the case of large number of constituents $p_{\mathrm{T}}D \rightarrow$ 0. Contrary to
the radial moment, gluon jets have a smaller $p_{\mathrm{T}}$D than quark
jets because their fragmentation is softer \cite{harvard}.

\noindent The use of these two shapes can help to discriminate between
two different physics scenarios: intra-jet broadening or collimation as a
result of jet quenching.  

\section{Jet reconstruction and corrections}
 \noindent Jets are reconstructed using the FastJet anti-$k_{T}$ algorithm
  with resolution parameter $R =$ 0.2, along with E-scheme
  recombination and using charged tracks reconstructed in the ALICE central barrel acceptance
  ($\eta <0.9$, $p_{T}>0.15$ GeV) assuming charged tracks to be pions.  The 10$\%$ most central Pb--Pb
  collisions were selected. 
The effects of the large underlying event in central Pb--Pb collisions
were addressed in two distinct steps, correcting separately for the
median background level and for its fluctuations. 

\subsection{Event-by-event average background subtraction}
\noindent The event-by-event estimate of the underlying event momentum 
and mass densities $\rho$ and $\rho_{\mathrm{m}}$ is done using a Fastjet
area based method\cite{Cacciari:2007fd}. This information is provided as input to two methods to subtract the background from the jet shapes:

\noindent-Area-derivatives method~\cite{Soyez:2012hv}: It involves a
  numerical determination of a given shape susceptibility to
  background and an extrapolation to zero background. 

\noindent-Constituent subtraction method \cite{Berta:2014eza}: It operates
  particle-by-particle so that the
four-momentum of the jet and its substructure are corrected simultaneously. 

\begin{figure}[h]
\centering
\includegraphics[width=0.52\textwidth]{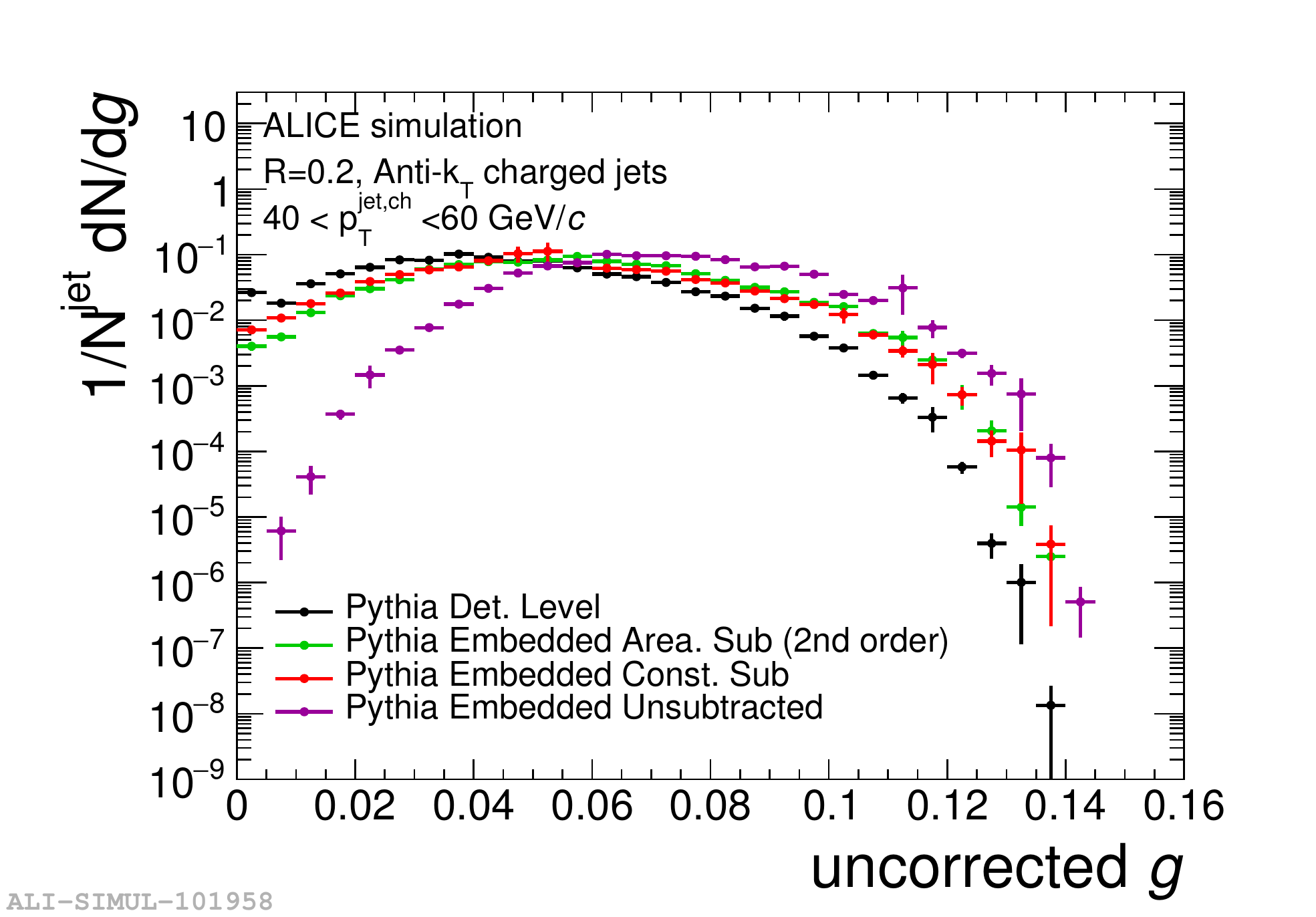}
\caption{Background subtraction performance}
\label{fig:bkgPbPb}
\end{figure}

\noindent The first method is used as default while the second is used to study
the systematic uncertainty due to method choice. 
\noindent To test the performance of the subtraction in Pb--Pb we embed PYTHIA\cite{Pythia} (Perugia11)
jets at detector level into Pb--Pb events. The results for the radial
moment are shown in Figure \ref{fig:bkgPbPb}. We compare the detector level
shapes (black symbols) to the subtracted hybrid shapes (red and
green) where by hybrid we refer to detector-level PYTHIA jets that are
embedded in Pb--Pb events, reconstructed and matched to the PYTHIA
detector-level probe.
 Residual differences between background corrected and detector level jets are due to
 background fluctuations and need to be unfolded. 
Note that no background correction is performed in proton-proton
collisions, where effects are negligible for $R = $0.2.

\subsection{Unfolding in two dimensions}
\noindent Residual background fluctuations and detector effects are
unfolded. We use Bayesian unfolding in two dimensions as implemented in RooUnfold
package \cite{RooUnfold} to obtain fully corrected jet shapes.

\noindent Since the unfolding procedure conserves counts, 
the raw input needs to be clean of combinatorial background for the sake
of stability of the procedure. The background response of jets, $\delta p_{\mathrm{T}}$, has a width of $\sigma=$ 4 GeV/c for
$R=0.2$ \cite{Abelev:2012ej}. The truncation of the raw yield at 30 GeV/c
sets our working point at more than 7 $\sigma$ away from zero and thus combinatorial background
is negligible. To unfold, we use a 4D response matrix with axes
(shape$^{\mathrm{part}}$,p$_{\mathrm{T,jet}}^{\mathrm{part}}$,shape$^{\mathrm{rec}}$,p$_{\mathrm{T,jet}}^{\mathrm{rec}}$).
Upper index 'part' refers to particle level and 'rec' refers to
reconstructed level. 
In pp, reconstructed level means detector level and the response is
filled using PYTHIA at particle level and after full detector smearing. 
In Pb--Pb, reconstructed level means detector level after correction
for the average background and smeared to account for fluctuations. To
construct the response matrix, we embed PYTHIA detector-level jets into Pb-Pb
events and we apply two successive matchings, between hybrid
and detector-level jets and between detector and particle-level
jets. 
\section{Jet shapes in pp}
\label{pp}
\begin{figure}[h]
\centering
\includegraphics[width=0.35\textwidth,height=0.32\textheight]{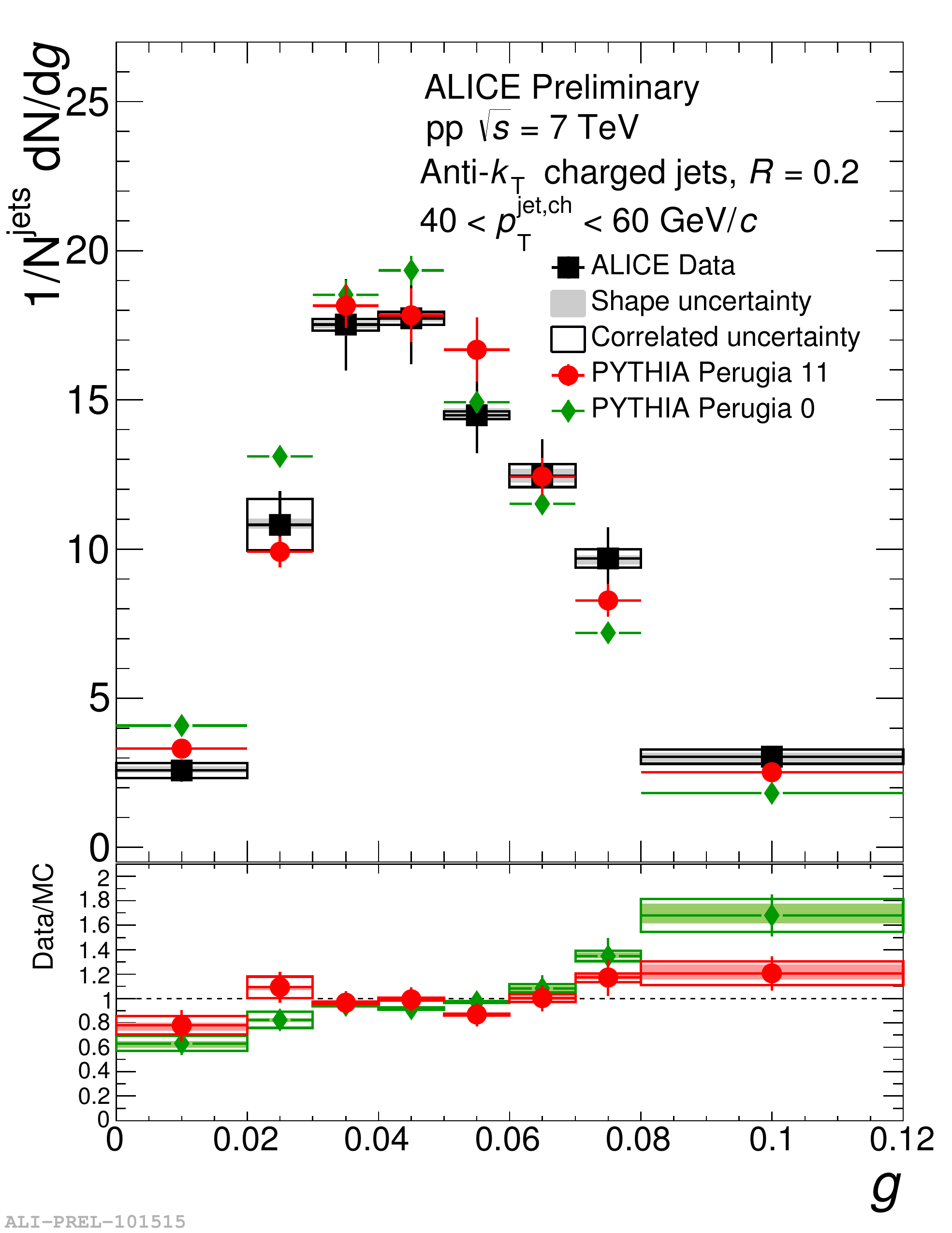}
\includegraphics[width=0.35\textwidth,height=0.32\textheight]{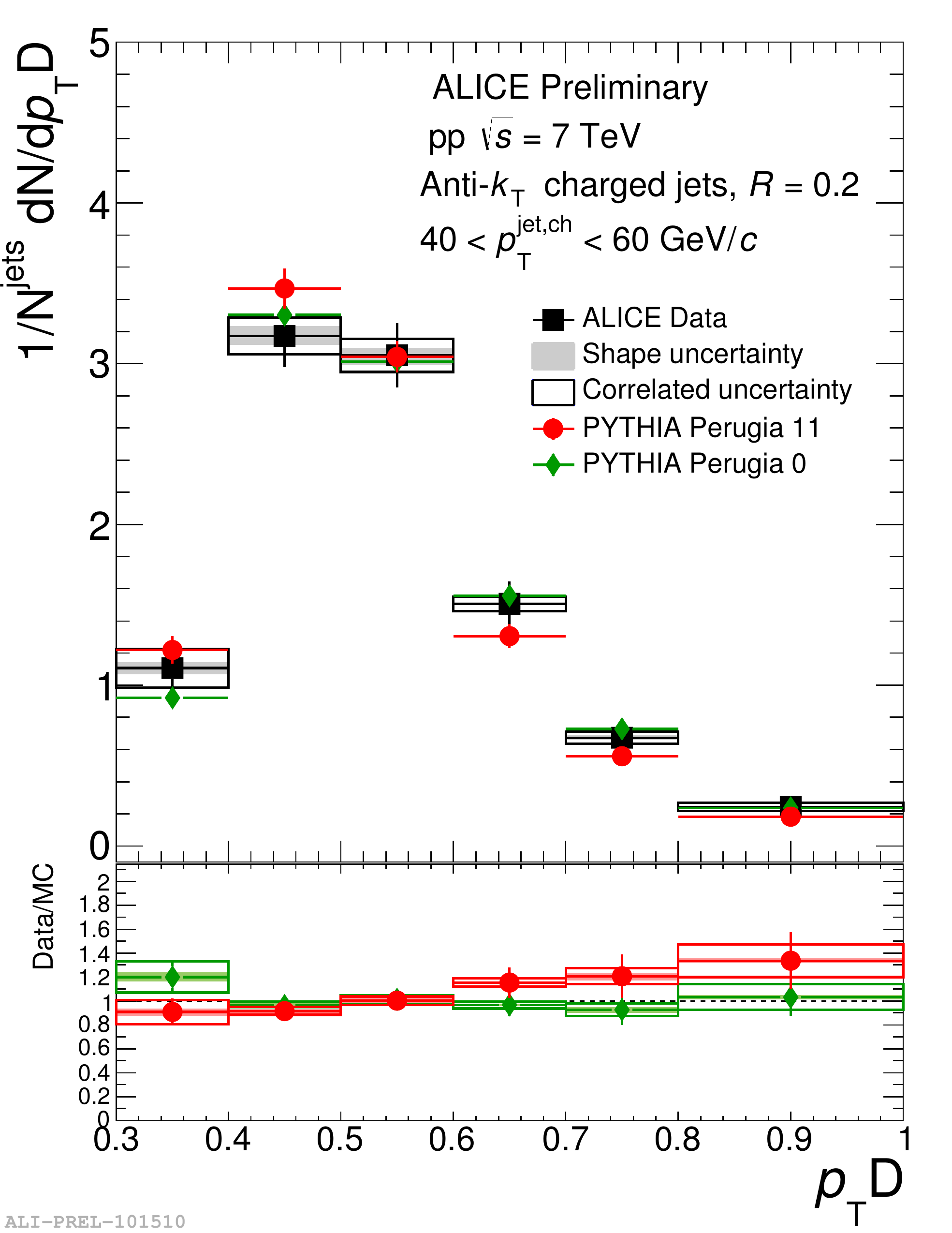}
\caption{Fully corrected shape distributions in pp for $R=0.2$}
\label{fig:RawDistpp}
\end{figure}

\noindent Figure \ref{fig:RawDistpp} shows the fully corrected shapes in
pp collisions at 7 TeV in the jet $p_{\mathrm{T}}$ range 40-60 GeV/c.  The results are compared to PYTHIA Perugia 0
and 11, which show a reasonable agreement given that non-perturbative effects are expected for small $R$. The systematic uncertainty is dominated by
single particle tracking efficiency uncertainty. Other sources of shape uncertainty are the regularization choice (we consider $\pm$3 iterations around default),
 the truncation value (we consider truncating the yield at $p_{\mathrm{T,jet}}$ 10 GeV/c lower than nominal) or the prior
 choice (we smear by 20$\%$ the prior correlation based by default on
 PYTHIA Perugia 0).   

\section{Jet shapes in Pb--Pb}
\label{PbPb}
\begin{figure}[h]
\centering
\includegraphics[width=0.44\textwidth]{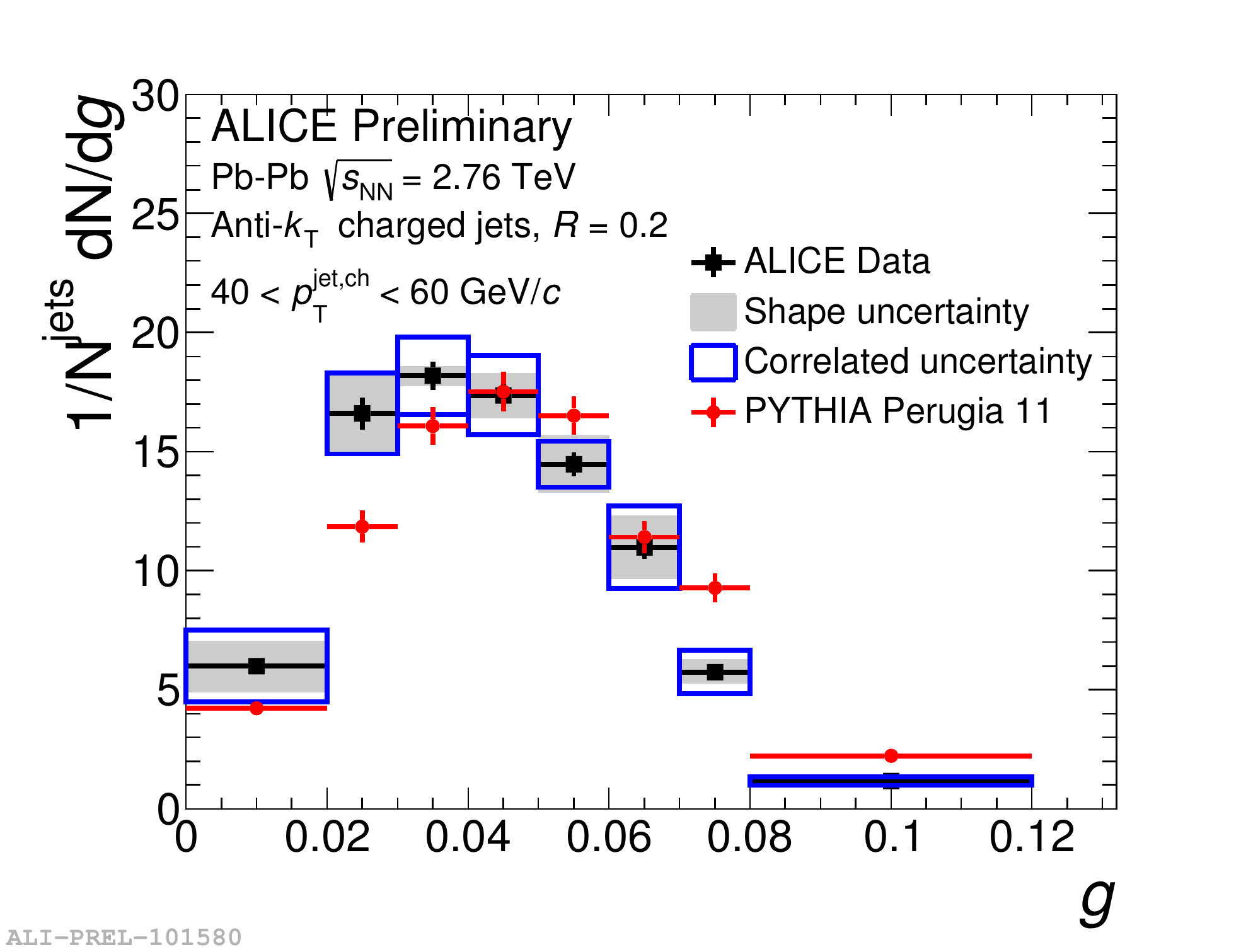}
\includegraphics[width=0.44\textwidth]{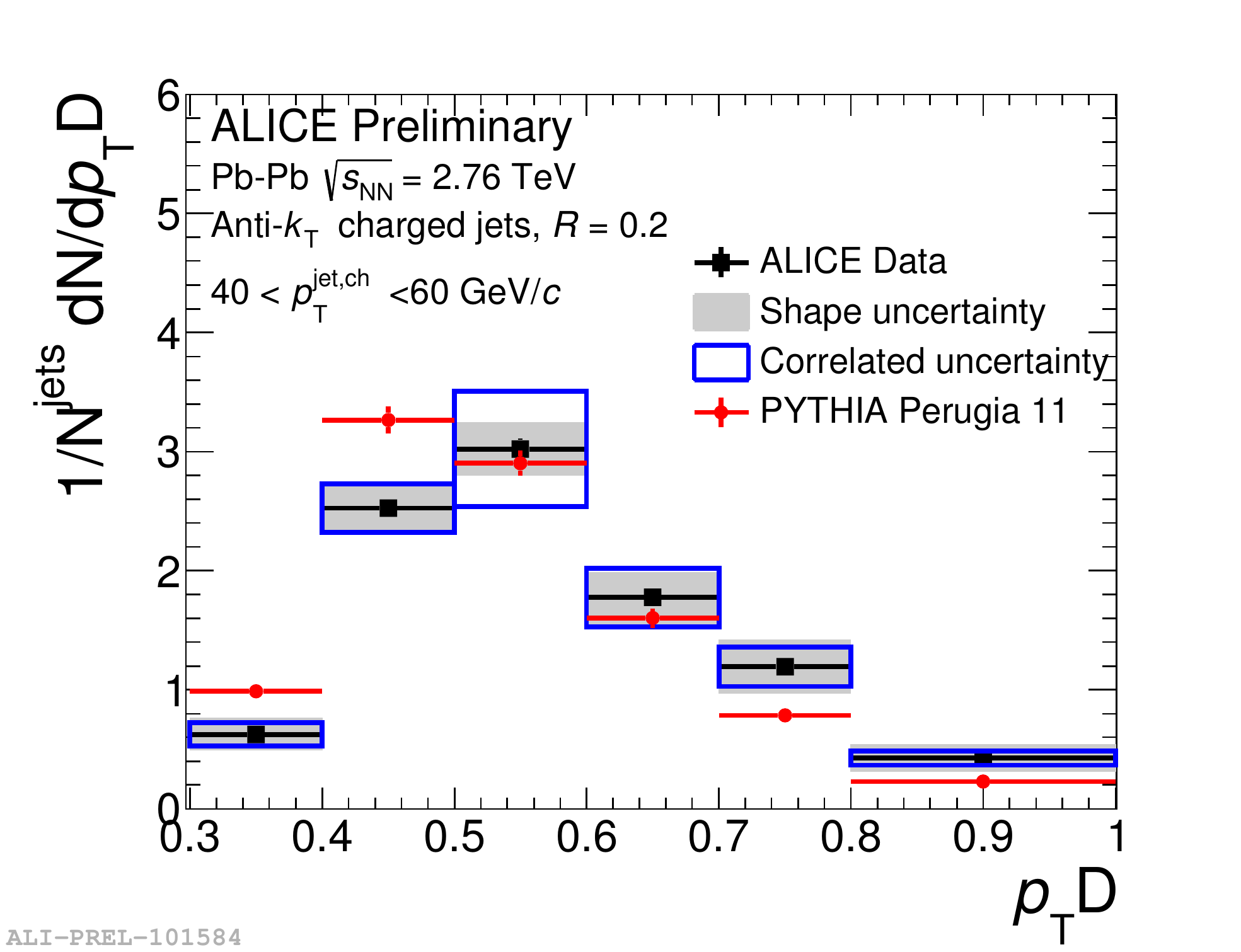}
\includegraphics[width=0.44\textwidth]{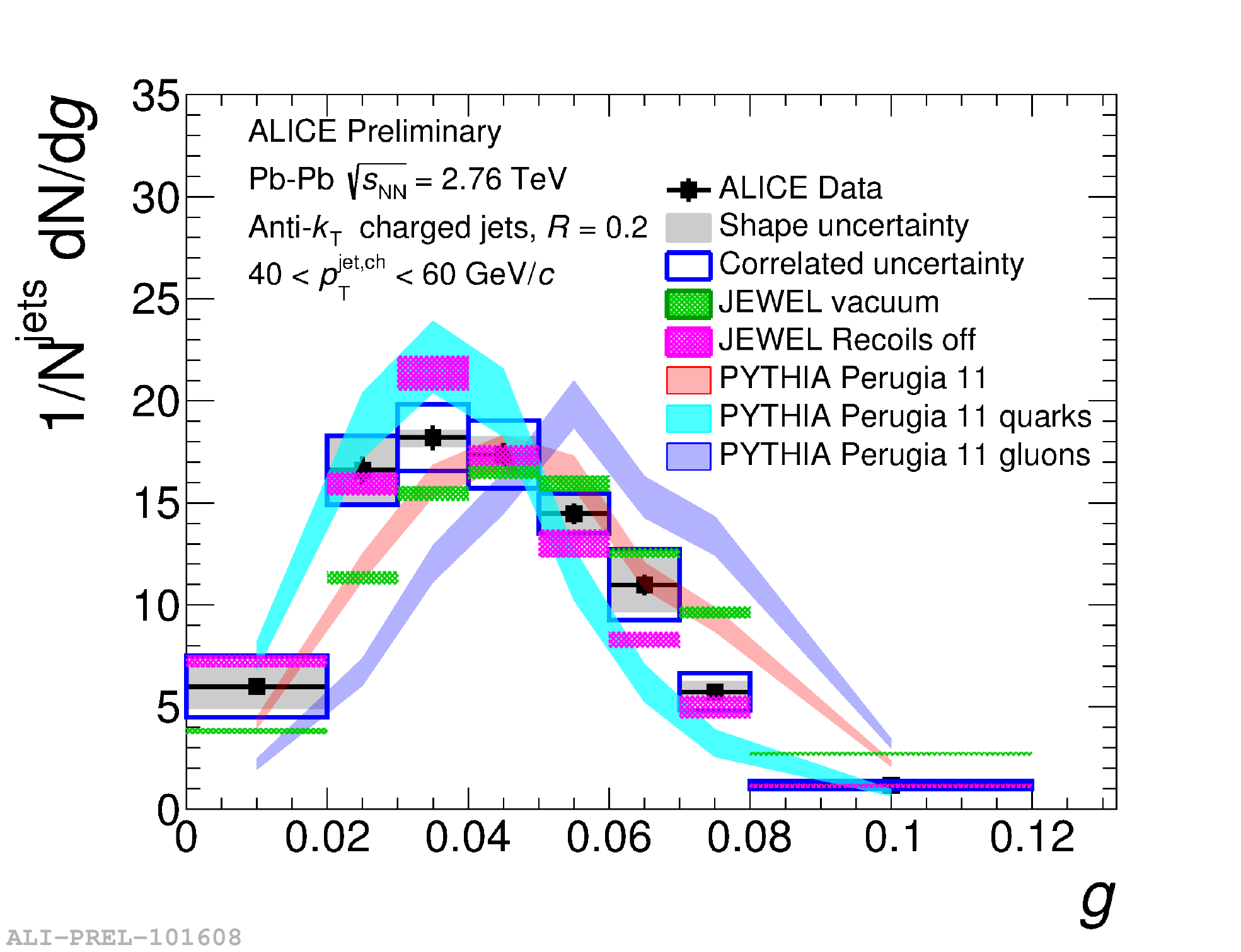}
\includegraphics[width=0.44\textwidth]{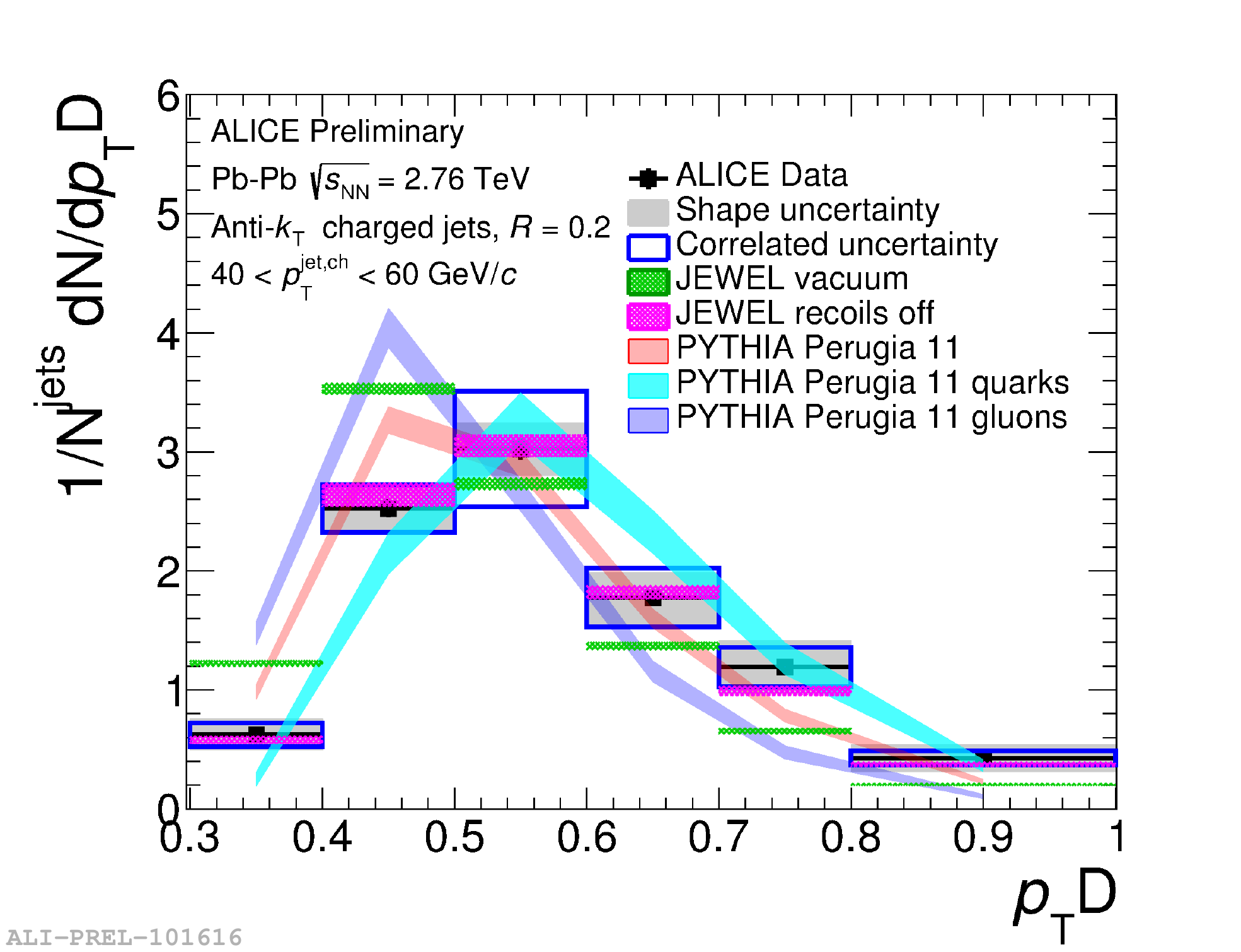}
\caption{Upper plots: Fully corrected shape distributions in Pb--Pb compared to
  PYTHIA Perugia 11. Lower plots: Fully corrected shape distributions in Pb--Pb compared to
  PYTHIA Perugia 11 inclusive, quark and gluon shapes and to JEWEL model}
\label{fig:RawDistPbPb}
\end{figure}

\noindent Upper plots in Figure \ref{fig:RawDistPbPb} show the fully corrected shapes in Pb-Pb
collisions at  $\ensuremath{\sqrt{s_{\mathrm{NN}}}} =$ 2.76 TeV compared to PYTHIA Perugia 11 in the
same jet $p_{\mathrm{T}}$ range of 40-60 GeV/c. Note that in addition to the systematic uncertainties considered in pp,
the background subtraction method choice contributes to the
shape uncertainty. The radial moment (upper left plot) is
shifted to lower values in data compared to PYTHIA. The $p_{\mathrm{T}}$D
(upper right plot) is shifted to higher values in data compared to
PYTHIA. Our results indicate that the jet cores in Pb--Pb are more
collimated and harder than the jet cores in PYTHIA at the same
energy. Due to jet quenching, when we compare jet shapes in Pb--Pb and pp at the same measured energy,
we might bias towards higher initial parton energy in Pb--Pb if a
significant fraction of the radiated energy is outside the used jet
cone.  Then the question is how the
energy was lost and how the radiation pattern of the jet was modified.
JEWEL\cite{Zapp:2013vla}  medium-modified jets are narrower and are harder than vacuum
jets at the same reconstructed energy \cite{marco}, in qualitative agreement with Pb--Pb
data as seen in Figure \ref{fig:RawDistPbPb}, lower plots.  The underlying physics mechanism in JEWEL leads to a collimation
of the jet, where soft modes are transported to large angle relative
to jet axis. For illustrative purposes quark and gluon vacuum
jets are added to the plot.  One can think of gluon jets as an
approximation to modified jets
in the hypothetical case where quenching accelerates the shower just
by increasing the number of splittings. This scenario would lead to a
broadening/softening of the in-cone shower (see differences in the
shape between inclusive jets and gluon jets in the plot) as
opposed to data.

\bibliographystyle{elsarticle-num}
\bibliography{<your-bib-database>}



\end{document}